# A2CI: A Cloud-based, Service-oriented Geospatial Cyberinfrastructure to Support Atmospheric Research

Wenwen Li, Hu Shao, Sizhe Wang, Xiran Zhou, Sheng Wu

GeoDa Center for Geospatial Analysis and Computation, School of Geographical Sciences and Urban Planning, Arizona State University, Tempe AZ 85287-5302

Abstract

Big earth science data offers the scientific community great opportunities. Many more studies at large-scales, over long-terms and at high resolution can now be conducted using the rich information collected by remote sensing satellites, ground-based sensor networks, and even social media input. However, the hundreds of terabytes of information collected and compiled on an hourly basis by NASA and other government agencies present a significant challenge for atmospheric scientists seeking to improve the understanding of the Earth atmospheric system. These challenges include effective discovery, organization, analysis and visualization of large amounts of data. This paper reports the outcomes of an NSF-funded project that developed a geospatial cyberinfrastructure—the A2CI (Atmospheric Analysis Cyberinfrastructure)—to support atmospheric research. We first introduce the service-oriented system framework then describe in detail the implementation of the data discovery module, data management module, data integration module, data analysis and visualization modules following the cloud computing principles—Data-as-a-Service, Software-as-a-Service, Platform-as-a-Service and Infrastructure-as-a-Service. We demonstrate the graphic user interface by performing an analysis between Sea Surface Temperature and the intensity of tropical storms in the North Atlantic and Pacific oceans. We expect this work to contribute to the technical advancement of cyberinfrastructure research as well as to the development of an online, collaborative scientific analysis system for atmospheric science.

Keywords: cyberinfrastructure, service-oriented, visualization, crawler, interoperability

1. Introduction

It is generally agreed that an atmospheric process refers to a systematic and dynamic energy change that involves physical, chemical and biological mechanisms. In recent years, atmospheric research has received increasing attention from environmental experts and the public because atmospheric phenomena such as El Nino, global warming, ozone depletion, and drought that may have negative effects on the Earth's climate and ecosystem are occurring more often (Walther et al. 2002; Karl and Trenberth 2003; Trenberth et al. 2014). In order to model the status quo and predict the trend of atmospheric phenomena and events, researchers need to retrieve data from various relevant domains, such as chemical components of aerosols and gases, the terrestrial surface, energy consumption, the hydrosphere, the biosphere, etc. (Schneider, 2006; Fowler et al., 2009; Guilyardi et al, 2009; Ramanathan et al., 2011; Katul et al., 2012). In complex earth system modeling, the data and services for atmospheric study present the characteristics of being distributed, collaborative and adaptive (Plale et al., 2006). The massive volume, rapid velocity and wide variety of data has led to a new era of atmospheric research that consists of accessing and integrating big data from distributed sources, conducting collaborative analysis in an interactive way, providing intelligent services for data management, and integration and visualization to foster discovery of hidden or new knowledge. One of the most important ways to support these activities is to establish a national or international spatial data infrastructure and geospatial cyberinfrastructure on which the data and computational resources can be easily shared, the spatial analysis tool can be executed on-the-fly and the scientific results can be effectively visualized (Yang et al., 2008; Li et al., 2011).

Technically, a geospatial cyberinfrastructure (GCI) is an architecture that effectively utilizes geo-referenced data to connect people, information and computers based on the standardized data access protocols, high speed internet, high-performance computing facilities (HPC) and service-oriented data management (Yang et al., 2010). Since the concept's official introduction by the National Science Foundation (NSF) in its 2003 blue ribbon report, cyberinfrastructure research has attracted much attention from the atmospheric science domain because of its promise of bringing paradigm change for future atmospheric research. As a result, several GCI portals to support atmospheric data analysis and integration have been developed:

- CASA (Collaborative Adaptive Sensing of the Atmosphere; http://www.casa.umass.edu), a real-time sensor network for acquiring and processing atmospheric data to predict hazardous weather events;
- LEAD (Linked Environments for Atmospheric Discovery; http://lead.ou.edu; Droegemeier, 2009), a service-oriented cyberinfrastructure that establishes a problem solving environment which hides complex computation details by providing efficient workflow management and data orchestration mechanisms;
- Polar Cyberinfrastructure (http://polar.geodacenter.org/gci; Li et al. 2015), the cross-boundary GCI that supports sustained polar science in terms of atmospheric, environmental and biological change.
- RCMED (Regional Climate Model Evaluation System; Mattmann et al., 2014), a cyberinfrastructure platform that supports the extraction, synthesis, and archive of disparate remote sensing data in various formats, including HDF (Fortner 1998), NetCDF (Rew and Davis 1990), Climate Forecast-CF (Eaton et al. 2003) etc.

These GCI solutions share similar goals and characteristics. They 1) provide precise and effective access to distributed geospatial data; 2) collaboratively analyze large-scale data through remote communication and sharing; 3) use an intelligent, systematic and user-friendly interface to improve user experience; and more importantly 4) support on-demand data integration, modeling and analysis. Achieving these goals simultaneously requires a new computing platform that supports scalable, elastic and service-oriented data analysis. Cloud computing platforms, known for their capability to provide a converged infrastructure and shared service to maximize the efficiency in data sharing, storage and computation, are a promising solution. First, the three levels of resource sharing strategies, Data-as-a-Service (DaaS), Software-as-a-Service (SaaS) and Platform-as-a-Service (PaaS), provide strong architectural and technical support that will enable the establishment of a service-oriented GCI for atmospheric research. Second, cloud computing can broaden the utilization of GCI by providing an open, collaborative, web-based environment for decision-making (Li et al. 2013). Finally, cloud computing techniques can provide strong support for efficient processing in a cyberinfrastructure environment in which many data/processing requests from distributed users may arrive simultaneously because of the cloud's ability to utilize the spare resources in a network of distributed computers (Yang et al. 2008; Yang et al. 2011). We believe that the integration of cloud-based GCI represents a new frontier for atmospheric research and that the marriage of the technology will foster the establishment of a software infrastructure that enables deep computing in atmospheric analysis and extensive use of cyber-technology across different communities and organizations.

In this paper, we describe our research in developing a cloud-based, service-oriented Atmosphere Analysis Cyberinfrastructure (A2CI) to support the effective discovery of distributed resources.

2. Literature review

In the literature, cloud computing has been widely used as the computing infrastructure to host various online decision making tools (Yang et al. 2011; Li et al. 2015) because of its advantages of being convenient, easy to manage and having an elastic capacity to handle various workloads over time. This popular service is also known as Infrastructure as a Service (IaaS), one of the four major types of cloud service defined by the National Institute of Standards and Technology (NIST). In comparison to IaaS, the other three types of cloud services—Data as a Service (DaaS), Software as a Service (SaaS), and Platform as a Service (PaaS), require a deeper level of understanding of the domain knowledge, including the distribution pattern of data, the computational characteristics of the analytical tools, and the extendibility of the cyberinfrastructure platform. We first provide a general review of the DaaS, SaaS, and PaaS principles, then we describe the design and development of cloud services following these principles to develop next generation cyberinfrastructure for atmospheric studies.

Data as a Service (DaaS): DaaS provides an effective way to make massive data readily available for clients and represents a great improvement over traditional approach. In conventional methods, accessing data to support atmospheric research requires the development of a standalone tool for each data repository. In the big data era, however, the countless number of data repositories, data structures, data formats, and other factors make it a significant challenge to build and integrate all kinds of customized tools for integrative data analysis. DaaS, which focuses on discovering needed data associated with a specific analysis task and delivering data to researchers without considering the physical location and sources of the data through the provision of services, becomes a promising strategy to address the aforementioned challenges. For data users, DaaS in GCI enables users from a wide range of communities to discover and access needed data by appropriate protocol. For data providers, DaaS reduces the cost of data storage and data management by moving the data server onto the cloud and increases data reuse and management efficiency through the provision of standardized data services.

Software as a Service (SaaS): SaaS provides on-demand tools and applications for performing specific tasks in the first layer on the top of a cloud server infrastructure. This elasticity enables atmosphere researchers to perform computing tasks without installing a model or analysis tool locally. For example, various analysis functions that monitor and predict atmospheric events can be deployed in a cloud environment and made available to online users. These functions can be deployed as OGC-compliant processing services, such as Web Processing Service (WPS), or Web Coverage Processing Service (WCPS) if a GCI primarily deals with raster data processing. Researchers access the online portal and send commands through the SaaS to run model simulations in the cloud. Some other services, such as data discovery service, data visualization service can also be encapsulated as services for easy integration into GCI solutions. The benefits of SaaS include eliminating local constraints of limited computing resources and software tools.

Platform as a Service (PaaS): PaaS enables virtualized computing between data providers and data users. In a cloud-based GCI, data and computing resources to satisfy the demands of atmospheric research can be provided using the mechanism of virtual space and virtual storage. Furthermore, the scalable data and computing resources defined on top of a cloud platform can be dynamically configured and scaled depending on the demand of atmospheric analysis request. In a PaaS model, it is not just the elements of CI, such as data or software provided, it is actually the entire CI problem-solving environment that is provided. In such an online environment, multi-disciplinary data is accessible through DaaS, the various software tools for atmospheric analysis are made available through SaaS, and the whole CI software platform is provided to support online data analysis and decision making.

The next sections describe the proposed CI framework for sharing, analyzing and visualizing atmospheric data in a cloud environment.

3. Cloud-based CI Framework for Atmospheric Research

Our proposed Atmosphere Analysis CyberInfrastructure (A2CI) was designed following the principle of service orientation (DaaS, SaaS and PaaS) in order to integrate as many data sources as possible and provide the best user experience. Figure 1 demonstrates the architecture of the A2CI.

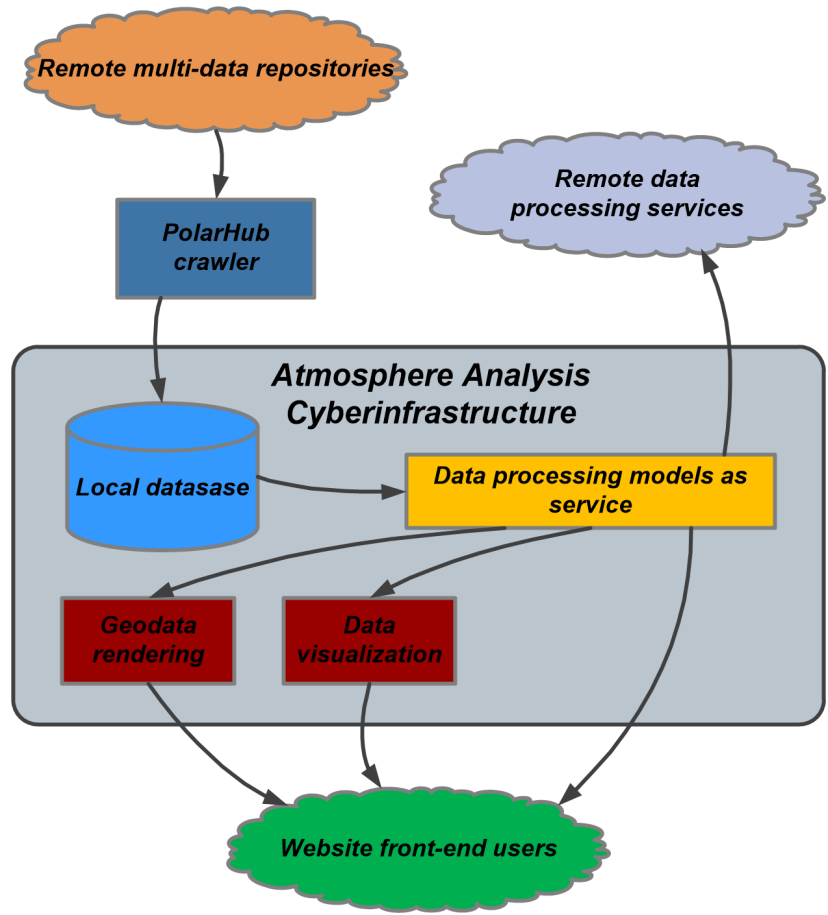

Figure 1. The architecture of A2CI.

There are five types of services being managed in the A2CI framework: data service, data discovery, data search, data processing, and data rendering and visualization. Atmospheric data is location-specific, and therefore, a typical type of geospatial data. These data are diverse and heterogeneous in nature. They have various encoding structures, i.e. vector or raster, due to different data providers' conventions. They may also have different spatial/temporal scales, spatial reference systems and accuracy, etc.

To ensure that data from various sources is sharable and interoperable, our A2CI implements *a data service module* that can handle disparate data according to Open Geospatial Consortium (OGC) standards. The OGC is an international industry consortium of over 500 universities, companies and government agencies participating in a consensus process to standardize publicly available web interfaces. This standard has been widely adopted by the atmospheric research community for sharing data and other resources. The data service module also provides the capability to publish raw datasets into services. To enrich the data repository in the A2CI system and support researchers' needs, a *data discovery service* is provided. This data discovery service is implemented through PolarHub, a large-scale web crawler that has the ability to identify distributed geospatial data resources on the Internet.

Besides managing data resources, a key feature of the A2CI framework is support for online spatial analysis of datasets. All the processing functions are developed as processing services such that they are easily transplanted and reused. Two types of processing services are enabled: *local data processing services* and *remote data processing services*. The local data processing services are the locally deployed spatial analytical tools and models as services. The remote data processing services refer to the analytical capabilities provided by another CI system with standard service interface. The capabilities of these services and their invocation protocol are stored in the A2CI database. Once needed, the high-level processing service will compose an analytical workflow that orchestrates the data and local or remote processing services for accomplishing an analysis task. Because of its ability to glue together other components and at the same time provide an array of useful functions to users, the "data processing models as service" becomes an A2CI core service.

The results generated, either in the form of a map or statistical chart, will be developed through the *rendering and visualization services*. The rendering service is mainly responsible for defining the symbology and style for rendering different types of geodata. It is actually part of the *visualization service*, which provides a platform that is either two-dimensional or multi-dimensional for intuitive presentation of the data and analysis results. The details of geodata rendering and visualization will be discussed further in Section 4.4.

All of these service components are coordinated by A2CI middleware to support a complete question-answering cycle from searching for data, integrating and mapping data, analyzing the data and visualizing the analytical results. These services are available as cloud services and are remotely programmable; hence they can be easily integrated and reused by other CI solutions to reduce code duplication and to accelerate the knowledge discovery process. The next section describes the service modules in detail.

4.     Components

4.1 PolarHub: a large-scale web crawler to provide data discovery service
Big data, which is dominating the Internet, has played an important role in improving our understanding of the Earth and supporting scientific analysis. There is an ever-increasing amount of earth observation data being recorded and made available online, including past and current atmospheric data. These data are widely dispersed on the Internet, making effective discovery a significant challenge. Various search engines, such as Google and Bing, are convenient for the public's daily information discovery needs. However, they are not suitable to search for the geospatial and GIS data that helps the ocean and atmospheric research. To bridge the gap, we have developed a web crawler, the PolarHub, and provided a data discovery service as part of A2CI to support scientific research, and effective and accurate access to the desired data.

PolarHub starts by visiting the Web from a set of predefined URLs then gradually follows the hyperlinks within these web pages to expand the search scope until the entire Web or most of the interlinked Web has been visited. While PolarHub is scanning web pages, it extracts any relevant geospatial data. As the crawling process continues, it identifies and downloads the metadata describing the content and the permanent link of the dataset. To efficiently accomplish the task, two questions need to be answered – 1) where do we start to visit the Web? 2) what kind of geospatial data is our target?

For the first question, we rely on search results from popular search engines as the crawling seeds. This is because the Web is like an ocean of data; but what we have is no more than a big fish net. Looking for data on the Web requires huge computing power to continuously crawl the web pages in it. Although general search engines such as Google and Bing are not effective enough to look for geospatial data, they are still utilizable here because they provide a great initial start for our spatial-data-oriented crawling due to the large number of web pages they have indexed.

For the second, the OGC Web Services (OWS) become the main data source for discovery because of the service-oriented principle adopted in our A2CI framework. Upon finding an OWS, PolarHub will invoke its semantic analysis module to determine whether a service is a related atmospheric dataset or not. The knowledge used in the semantic analysis is a bag of words imported from NASA GCMD (Global Change Master Directory) science keywords. These words are scientific terms used to measure the atmospheric condition, such as "Albedo" or "Atmospheric pressure measurement". The services not related to atmospheric science are disregarded by PolarHub. Figure 2 describes the software framework of PolarHub, which is composed of the crawler component and the storage component and interacts with the A2CI data management component.

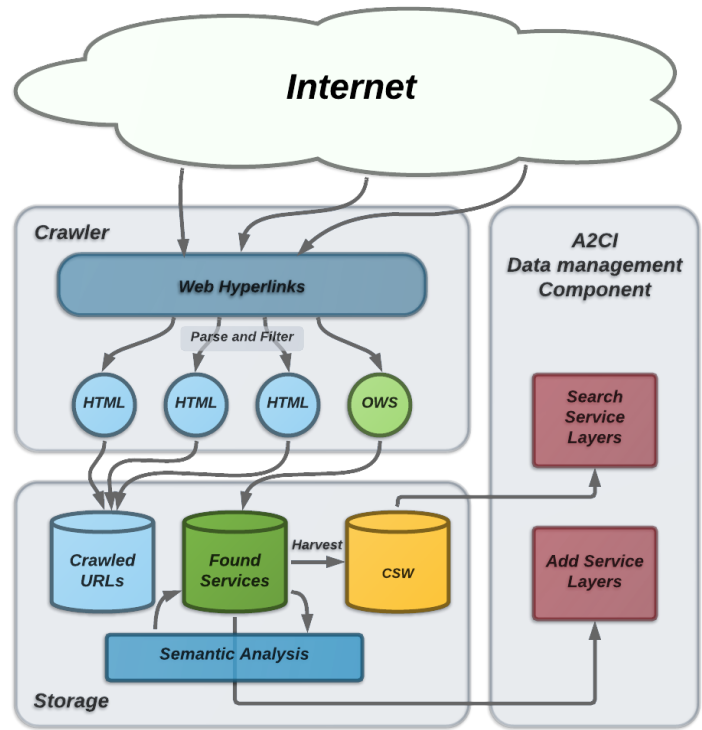

Figure 2. System architecture of PolarHub

The crawler component is responsible for crawling the Internet and finding relevant data. A typical crawling task starts with keywords that indicate the theme of the desired geospatial data. Taking the search for atmospheric data as an example, a user may enter "World Sea Surface Temperature" or simply "World SST" as a start. Once it receives the keywords, PolarHub will redirect the query to a general search engine such as Google or Bing to retrieve a set of relevant web pages as crawling seeds. All of these seeds will be put into a seed pool that holds the URLs waiting for processing. There are mainly two kinds of URL processing. One is for URLs that link to an HTML webpage. The other is for the XML (Extensible Markup Language) documents, in which the OWS metadata is encoded.

At the beginning of processing, the crawling engine retrieves a URL from the seed pool then analyzes the file type of the URL, i.e., whether it is an HTML webpage or an XML metadata file. The extracted URLs linking to a dynamic HTML webpage are then inserted into another seed pool. At the same time, these URLs are checked to determine whether they belong to a specific type of OWS. If so, the link information together with all the metadata associated with the data or processing service will be sent to the local data repository for caching.

To foster cross-CI data discovery, namely to make PolarHub discovered data available to other CI portals, all the local data will be harvested into an OGC-compliant online catalogue with a CSW (Catalogue Service for the Web) interface openly accessible by a web client or other portals. This catalogue supports full-text searching through a structured query language, Contextual Query Language (CQL), and interfaces with A2CI middleware's data management component. This provides an important data source for atmospheric data search and analysis.

Taking advantage of the data discovery service provided by PolarHub, A2CI has successfully discovered 1,816 atmosphere-related datasets. These datasets are distributed across 32 countries including the US, Canada, Australia and several EU countries. Figure 3 demonstrates the spatial distribution of the web servers that host these datasets.

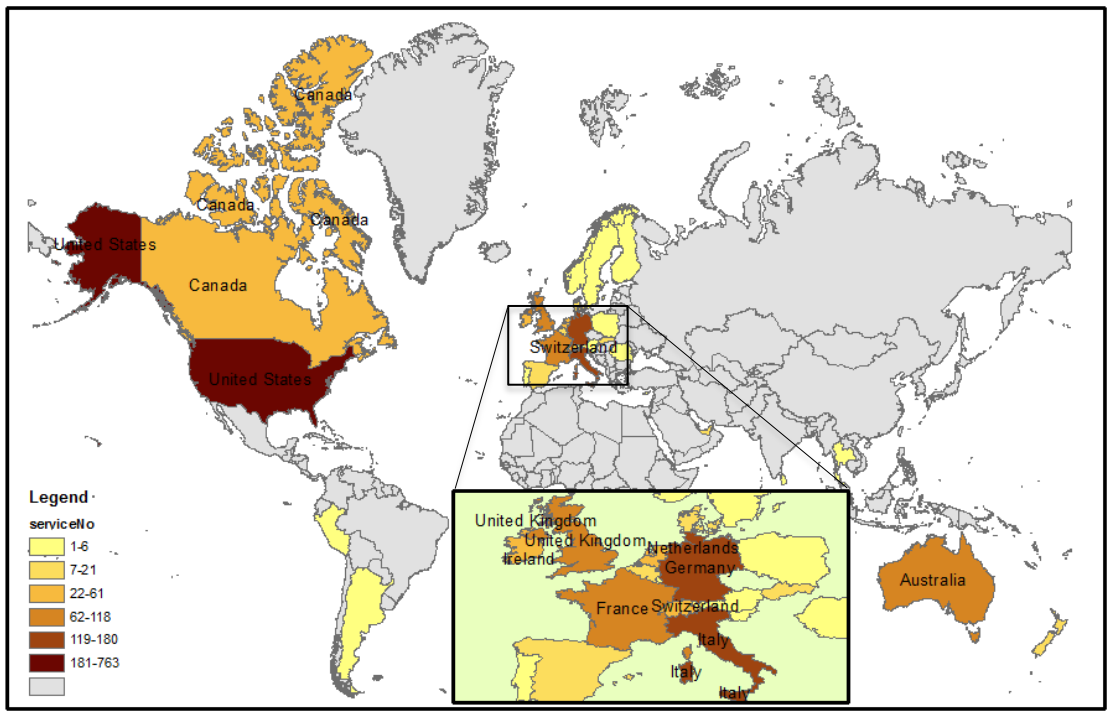

Figure 3. Spatial distribution of web server hosting atmospheric related data per country. Six categories are identified and "Natural Break" is used as the classification methods. Grey color means no data.

The labeled countries on the main map are those providing more than 62 atmospheric data services (top three classes per natural break classification). The inset map shows a closer look at the service providers located in Europe. Among all these countries, the US provides almost half of the total datasets in the OWS format. European countries such as Germany, Italy, UK, as well as Canada and Australia are all main providers for the atmospheric datasets. None of the relevant data are found in Asian countries, such as China, and or in any of the African countries, such as Libya and Sudan. This pattern reflects the fact that western developed countries are more active in sharing data as services for reuse and integration. We did not find any OGC related data from Asian developed countries such as Japan or Korea. This may due to the relatively less active role of these countries in the OGC community.

In the US, Federal government agencies that focus on earth and space research, including the USGS, NOAA, NASA and NationalAtlas.gov, are the main contributors for atmospheric data services. Besides these government agencies, several US universities and national data centers, such as the University of Hawaii and National Snow and Ice Data Center (NSIDC), are also contributing to the sharing of atmospheric data. Figure 4 lists the top 10 atmospheric data providers in the US as determined through a further analysis on the atmospheric data holdings existing in the A2CI data portal. The available datasets provide us rich resources to conduct atmospheric analysis. At the same time, however, an effective data/service management mechanism is needed for better coordination of these distributed data resources. A2CI aims at providing a flexible and easy-to-use platform to support this coordination.

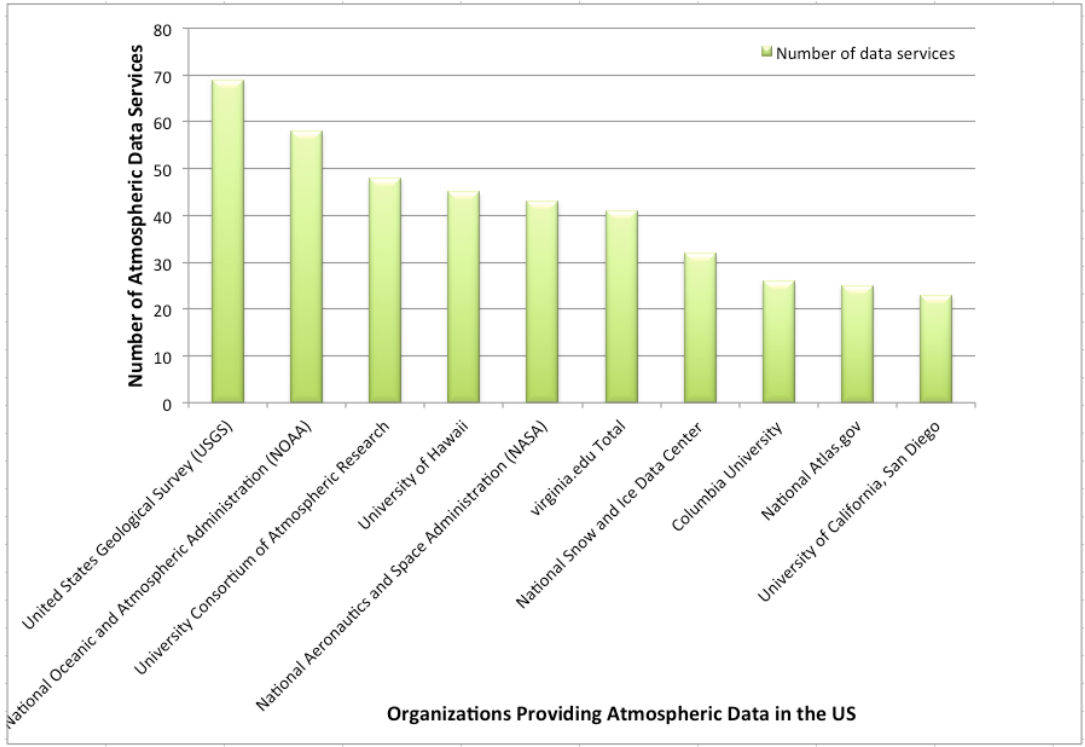

Figure 4. US organizations providing atmospheric data services

4.2 Data and service management

The A2CI platform is user-centric as different users can work on different applications or analyze multiple data of interest. All data used for a specific application is organized into a workspace. One user can create multiple workspaces, meaning that they can start multiple research topics on our A2CI platform. All user-created data and intermediate results are saved in the A2CI database to facilitate multi-phase scientific analysis. To support this, we developed a data and service management module to manage user data, workspace data and atmospheric data collected by PolarHub as well as other analysis-related information. Figure 5 demonstrates the UML diagram of the database schema.

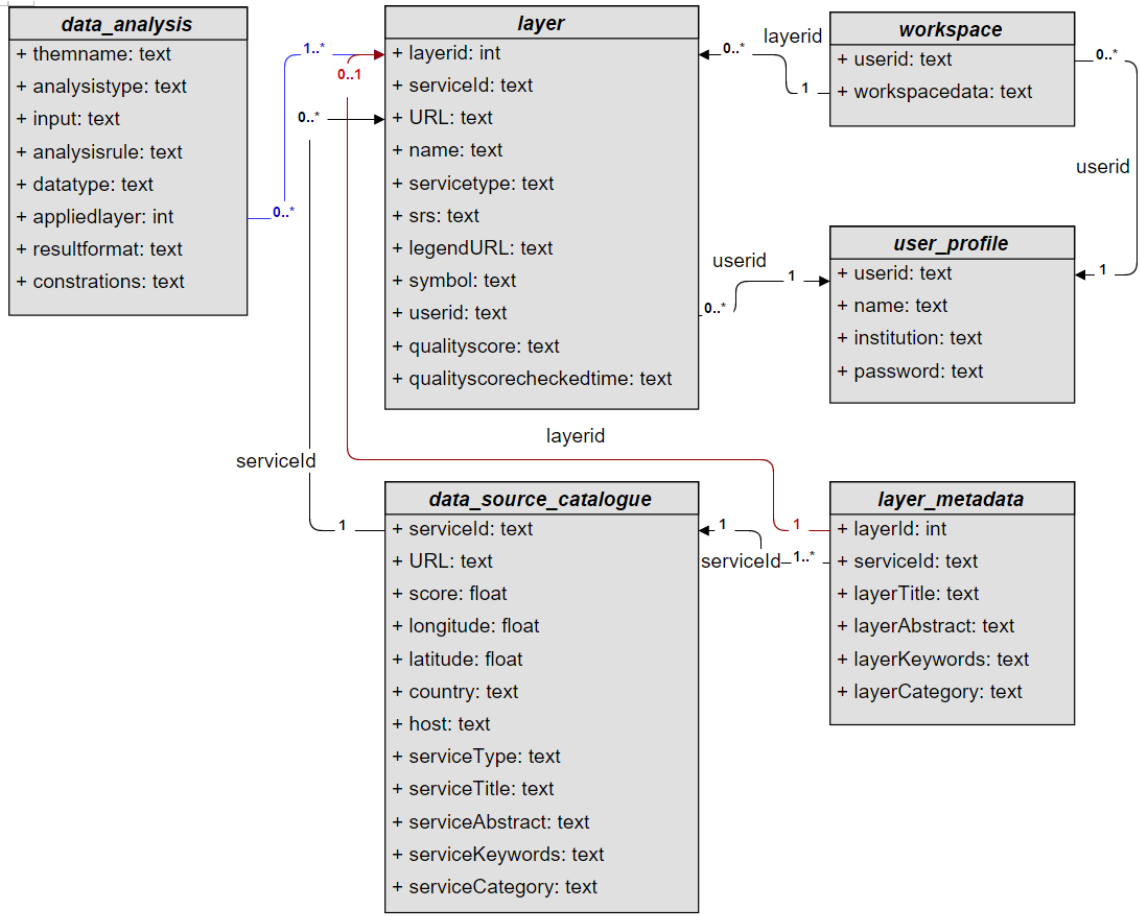

Figure 5. UML diagram of A2CI database schema

The *user_profile* table records users' email addresses as their unique ids and other profile information such as full name, institution, etc. This information is obtained during the registration process. User's passwords are stored as an encrypted string. The user id also plays a key role for linking different records together among the "*user_profile*", "*layer*" and "*workspace*" schemas. Any end user who wants to conduct analysis on the A2CI portal must register in the system and maintain a user profile. This way, the user's data viewing and analysis history can be retained and users' working results can also be saved for multi-logon processing.

The *workspace* table contains all of the essential attributes of a user-created workspace, including the workspace name, the default spatial reference system used, all data layers in a workspace, the layers' display order, etc. This information is formatted as a JSON string and compressed into one single data column titled *workspacedata*. Each time a user logs into the A2CI portal, the workspace information will be directly loaded into the web front-end and processed by JavaScript. Since JSON strings were originally designed for the JavaScript language, saving workspace data into JSON makes it much easier for the front-end web portal to present and visualize workspace data. The *workspace* table is linked to the *user_profile* table by the unique user id. The data layers loaded in each workspace are linked to the layer table by unique layer ids.

Different workspaces can load the same data layer, as long as this information is explicitly listed in the *workspacedata* field in the *workspace* table. When a user deletes a data layer in a workspace, the layer record will not be deleted in the layer table, instead only the record linking the workspace and the data layer will be removed. This way, different users' or a single user's operations on the same data layer in different workspaces will not be affected. Moreover, since one data layer only has one permanent record in the database no matter how many times it is used by various workspaces, data redundancy can be avoided.

The layer information is stored in two data tables: the *layer* table for storing visualization and analysis-related data used in the A2CI system and the *layer_metadata* table for storing the static layer information associated with the data service it belongs to. In the *layer* table, essential information such as layer name, its web URL, and spatial reference information are saved. When a data layer is requested by a user in the A2CI portal, this information will be encoded into a JSON object and pushed to an OpenLayers' web client to generate a data retrieval request, download remote data and display it in the map container. The *userid* records the user that the layer belongs to. The *symbol* column records the customized symbol schema for each layer, such as opacity, size and color of the layer's features. The *quality score* of each layer indicates the service's performance of each layer, such as the data transmission rate. This *layer* table is linked through the primary key *layerId* with the *layer_metadata* table, which stores additional thematic information about the layer's content, such as the title, abstract and keywords. Though this information will not be used for data retrieval, they are important in supporting the discovery of relevant datasets by matching the thematic data.

These two layer tables are mainly responsible for managing data layers that are the atomic unit for visualization. To manage the data resources at a higher level – the data service level, we designed another table, the *data_source_catalogue*, as the data layers come from different data services. In fact, PolarHub uses this data table to manage all data services crawled. The crawled data service information includes the URL, which links to the original data provider, and the physical locations of the service hosts in the *latitude, longitude* and *country* columns. The *score* is a quantitative measure to indicate the quality of the data service. The service table and the two layer tables are linked by the unique *serviceId*. A layer discovering service is designed and implemented to help users efficiently find appropriate data resources on a layer level by searching from the *layer_metadata* table or at a service level by redirecting the search to the *data_source_category* table.

Besides the mechanism to manage data resources, another database table, *data_analysis*, is also included. Many data analysis models are implemented as services in A2CI's front-end and back-end. The profile information of these analysis services, including service name, service URL, the input/output of each service model, data type that each service takes and generates, analysis rule description, and the special constraints of each model, as well as the data layer in the workspace that binds and presents the results are all stored in this table. Using this information, the A2CI front end can generate a scientific workflow automatically by linking the data and analytical components, significantly accelerating the process of data analysis and knowledge discovery.

4.3 Service-oriented data integration

The conventional method of data integration accesses all atmospheric-study-related data sources and stores them in a local data repository. The countless number of data repositories, data structures, data formats, and other factors in the big data era make it impossible to build a local database and integrate all the independent data repositories together. Therefore, a service-oriented data integration strategy is adopted in A2CI. It was designed following the principle of DaaS to provide an effective way of making massive data available for clients without the need for a local database. To save storage space and reduce management efforts only the metadata about a distributed dataset is recorded. When users require the actual data for viewing or analyzing, the data will be directly retrieved from the original data provider and transferred to the end users. This data integration strategy significantly reduces the data loading pressure on the server end of the A2CI platform. By crawling as many atmospheric-study-related data sources as possible and providing a comprehensive data search interface to users, A2CI can also increase users' data discoverability and usability. Figure 6 represents the workflow of the service-oriented data integration in A2CI.

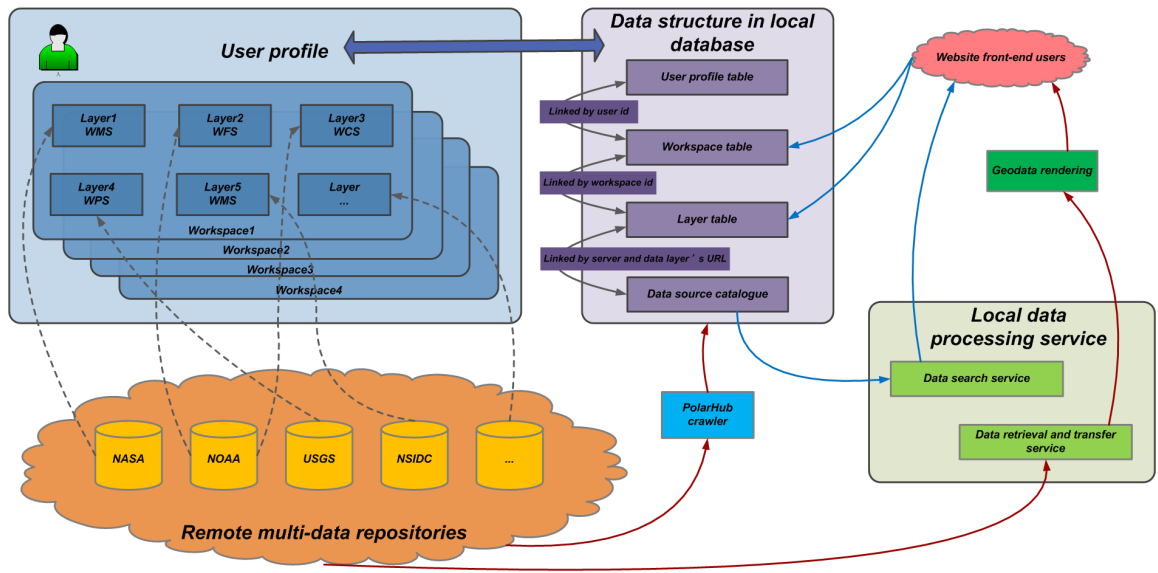

Figure 6. The workflow of service-oriented data integration

The first level of integration occurs for presenting and overlaying disparate data sources. As discussed in the previous section, these datasets could come from remote data repository hosted by government agencies such as NASA, NOAA, USGS, or from some large data centers such as NSIDC. Through a data service layer, these data could be encoded into different OWS formats according to different application needs. For instance, remote sensing images in raster format could be published into the OGC Web Map Service (WMS) for visualization. Using this service standard, the backend datasets, regardless whether they are in GeoTiff or GRID formats, will be rendered into a static map covering the requested geographical region. The static map if requested and returned in PNG format, can be tailored to set the transparency level at the client side to make a better visual effect when overlaying with other data layers.

Raster-based atmospheric data can also be shared in the OGC Web Coverage Service (WCS), which returns actual data in comparison to the static image given by a WMS. The disadvantage of a WCS is that the data file is much larger than the size of the static image. Vector-based data, in comparison, can be shared through the OGC Web Feature Service (WFS), which serializes the raw data formats, in an ESRI shapefile for example, to a standardized GML (Geography Markup Language) or Geo-JSON (JavaScript Object Notation) for client visualization. These feature data can be customized with different display styles, including the use of different color ramp, point or line size, and color to enhance the visual analysis in combination with raster datasets, etc.

Note that a successful integration of disparate data for conducting atmospheric analysis requires an intelligent method to search for the most suitable dataset from a set of data holdings. To accomplish this, we developed a data search service. Different from the data discovery service, which focuses on locating the distributed resources from the Internet to build a large atmospheric data clearinghouse, the data search service deals with finding relevant dataset that satisfies a user's request from the clearinghouse. This service provides a comprehensive search interface for users to find the data resources meeting certain conditions such as keywords, time period, formats, region of interest and spatial reference system (SRS). The search results will be presented to users as a formatted table. Additional functions are also provided to help the user better understand the data source. For example, a data quality score regarding the server response time, degree of metadata completeness and other factors will be calculated and reported to the user. A preview thumbnail image of the data will also be provided if a standard OGC WMS or WFS protocol publishes the data. Once users select a data source, the data source will be stored in the user database as a record associated with their user profile. This way, even if the data sources are distributed in many different remote data repositories and vary a lot, they can be uniformly integrated and stored in A2CI's local database.

The data search and integration strategies facilitate the integrated analysis of data from multiple sources. This integration is at a more static level, which means visual analysis of atmospheric phenomena is based upon the information that already exists in the datasets. In order to realize integration at a higher knowledge level, analytical tools are enabled in the A2CI platform as OGC services – the OGC Web Processing Services (WPS). These WPSs take the static data as input and create new information and knowledge through different analysis methods. This type of service-orientation belongs to a specific type of SaaS (Software as a Service) - Spatial Analysis as a Service (SaaaS) in the cloud-computing paradigm. The advantage of publishing various spatial analysis tools, such as zonal statistics, into web services, is to increase their reusability and enhance A2CI's analytical capability by fusing existing data and creating unknown knowledge to further advance science.

4.4 Multi-dimensional visualization service

Once the data and analytical results are generated through the cloud data processing services, they are sent to the A2CI's visualization service for mapping and rendering. Both 2D and 3D visualization mechanisms are provided. The layer data management module controls the visualization mode and all the visualization-related data, including all the layer information, their display orders, as well as the uniform projection data. When a user chooses to switch between the 2D and 3D visualization, all the layer information is sent to the corresponding visualization engine for result rendering. Figure 7 presents the data visualization framework.

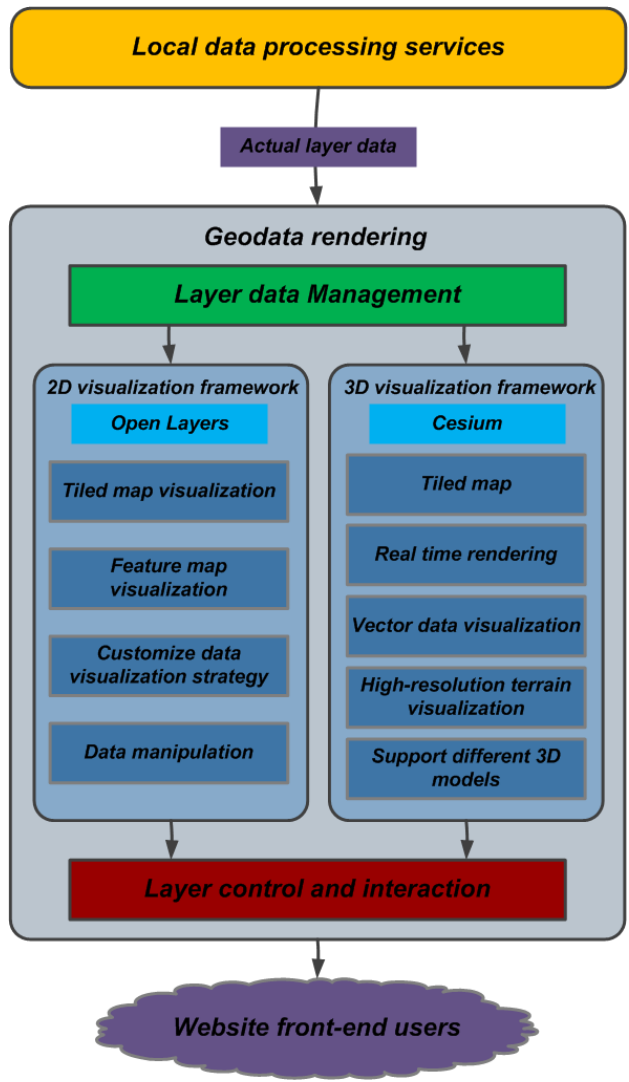

Figure 7. Multi-dimensional visualization service

**2D visualization service**

The 2D data rendering service is built based on OpenLayers, an open-source geo-data visualization and manipulation library in JavaScript. The 2D mapping module includes the following main functions to visualize atmospheric dataset:

(1) Use of OpenLayers in A2CI portal for image and feature map visualization: OpenLayers supports standard WMS protocol and is able to import different kind of image formats such as TIFF, JPG, and PNG into its map panel. In addition,

OpenLayers also provides a built-in interface for integrating base maps provided by commercial companies, such as Google, Esri, Open Street and Bing, which helps users better understand the context of their study region. OpenLayers is also capable of importing feature layers in various formats, such as Esri JSON, GeoJSON, GML, KML (Keyhole Markup Language), WFS, and XML files. What's more, OpenLayers also supports basic geofeature re-projection between different SRS (Spatial Reference System), which means OpenLayers makes users' available layer source much richer.

(2) Tiled map visualization: one of the main data sources in A2CI is OGC WMS, which returns static image from the remote server. When the requested extent becomes too large, loading and visualizing the entire dataset as a single piece may become very slow. Tile-based visualization resolves this issue by partitioning big datasets into smaller pieces with each piece's data individually and concurrently requested, retrieved, and displayed. This parallel processing strategy helps improve system performance and user experience.

(3) Customize data visualization strategy: In the A2CI portal, users can define the opacity of tiled and feature maps. For the feature maps, users can also customize the symbols of layers, such as fill color, stroke color of polygons, and symbol size of points, presenting rich layer attributes to users.

(4) Data manipulation: Through the powerful interaction interface provided by OpenLayers and A2CI visualization service, users can draw their own features or revise existing feature data on the 2D map.

**3D visualization service**

Besides the 2D map container, A2CI also supports the visualization of atmospheric data in a 3D globe. As we know, 2D maps allow us to view the data on the entire Earth in one scene, but 3D maps usually cannot. However, 3D visualization has the advantage of introducing little or no distortion in the original data, in comparison to a 2D map. Therefore, both visualization strategies are supported as they can compliment each other to realize the best visualization effect for presenting atmospheric data in the A2CI portal.

The embedded 3D virtual globe is built upon Cesium, an open-source virtual globe made with WebGL technology. This technique utilizes graphic resources at the client side by using a JavaScript-based library and WebGL to accelerate client-side visualization. The virtual globe has the capability of representing many different views of the geospatial features on the surface or in the atmosphere of the Earth, and can support the exploration of a variety of geospatial data. By running on a web browser and integrating distributed geospatial services worldwide, the virtual globe provides an effective channel to find interesting meteorological phenomena or the correlation between heterogeneous datasets (Hecher et al. 2015).

As part of the visualization service, the virtual globe can dynamically load and visualize different kinds of geospatial data, including tiled maps, raster maps, vector data, high-resolution worldwide terrain data and 3D models. Below we list the main functions supported by the 3D visualization module:

(1) The world bathymetry and other terrain data is georeferenced and pre-rendered to serve as the base map of the virtual globe. All of the tiled map services, such as Web Map Tile Service (WMTS) specification developed by the OGC, Tile Map Service specification developed by the Open Source Geospatial Foundation, Esri ArcGIS Map Server imagery service, OpenStreetMap, MapBox and Bing maps, can be easily loaded into the virtual globe as base map.

(2) Besides the base map, the virtual globe can accept real-time rendered WMS map services and geo-referenced static raster maps as imager layers to overlap on the base map. All the imagery layers can be ordered and blended together. Each layer's brightness, contrast, gamma, hue, and saturation can be customized by the end user and dynamically changed.

(3) The virtual globe also provides a library of primitives, such as point, line, polygon, rectangle, ellipsoid, etc. for vector data visualization. In order to improve the transmission speed of big atmospheric data over the Internet (Li et al. 2015), open specifications such as GeoJSON and TopoJSON are adopted to encode collections of simple spatial features along with their non-spatial features.

Besides these primitives, the virtual globe visualization module also supports complex spatial entities by aggregating related visualization and data information into a unified data structure.

The A2CI will provide uniform layer control and interaction interface on both 2D and 3D visualization frameworks to help users efficiently manipulate and customize their data layer and to conduct experiments.

5   Graphical User Interface of A2CI

Figure 8 demonstrates the GUI of the A2CI portal, which as a whole can be considered as a PaaS – Platform as a Service, and is deployed onto an Amazon EC2 cloud infrastructure.

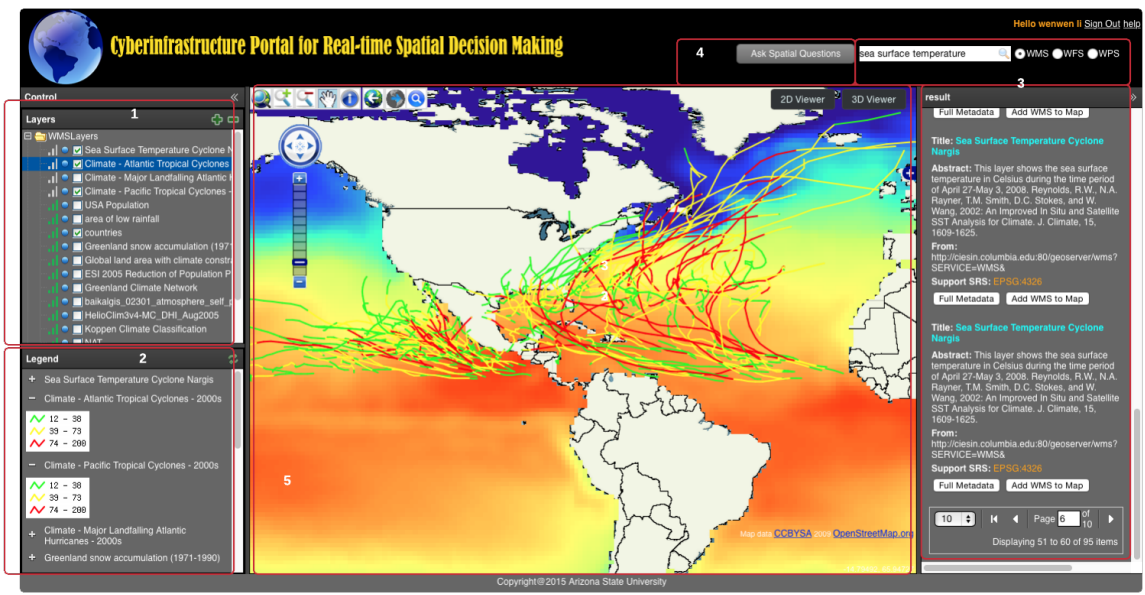

Figure 8. Graphic User Interface (GUI) of A2CI portal

Box 1 (top left navigation) lists the layer management module, in which a list of data layers is included. These data are loaded from the database once the user is recognized in the system. Box 2 (bottom left navigation) is the legend component that displays the information of all visible layers. This data is loaded in real-time from a remote data server at the same time the map data is requested. Box 3 (top right navigation) is the data search service, supported by the OGC CSW (Catalogue Service for Web) 2.0.2. The small box at the bottom shows the results returned by searching data related to Sea Surface Temperature (SST). An abstract of each resulting data layer is provided. Users can use this information to decide whether the data is of interest. If so, the user can click the "Add WMS to Map" button to add the service to the map panel for visualization. Box 4 (right header) is the module for various spatial analyses. For instance, zonal statistics of the SST can be conducted on a state basis to obtain a long-term mean temperature for different areas. Box 5 (middle) is the map service. In the example, four data layers from three remote servers (the nationalatlas.gov, ciesin.columbia.edu, nsidc.org) are integrated for atmospheric analysis. The line data shows the trajectories of all tropical storms in the 2000s in the Pacific and Atlantic Ocean, respectively. Different colors represent different frequencies on a particular trajectory. The blue-red image data shows the SST data in 2008. The country boundary data from NSIDC is also loaded.

The same data can also be loaded onto and visualized in the 3D globe, shown in Figure 9. From the integrated map, we can see that tropical storms originating in the Atlantic Ocean travel longer distances and have a higher intensity than those developed in the Pacific Ocean. This means that Atlantic Ocean storms carry more energy than Pacific Ocean storms as they travel. The energy could potentially come from the warmer weather in the same latitude in Atlantic Ocean than the Pacific Ocean, reflected in the SST data in the 2D map. Further analysis on the intensity can reveal more quantitative evidence on atmospheric conditions and the formation of tropical storms.

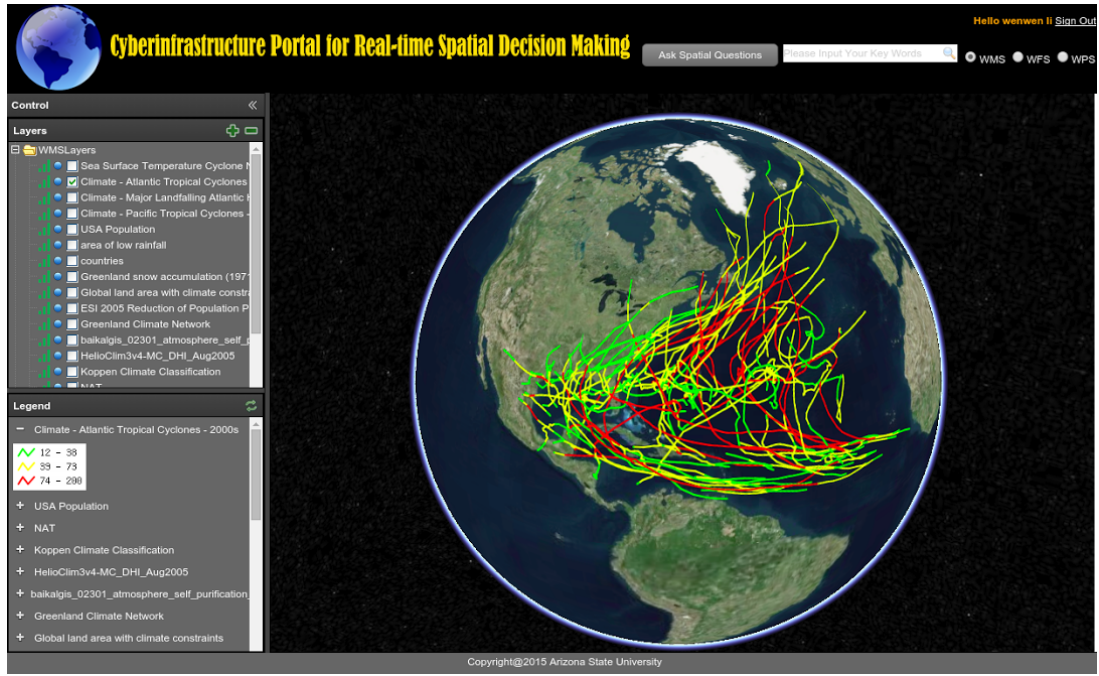

Figure 9. A2CI 3D visualization

6   Conclusion and discussion

This paper introduces a service-oriented cyberinfrastructure, the A2CI, to support atmospheric research. This CI framework introduces and integrates four primary principles—DaaS, SaaS, PaaS and IaaS—in the cloud-computing paradigm. There are a number of advantages of adopting these design principles. First, the use of cloud-computing platforms, such as Amazon EC2, to deploy the whole A2CI platform (the IaaS principle) can eliminate the hurdle of managing web servers and at the same time achieve high stability of the operational system. Second, the highly modularized, data service and software service-oriented design (following the DaaS and SaaS principles) makes the software system and its data and analysis resources highly extendable, sharable and reusable. Third, as web-based analysis is becoming more common, sharing the A2CI framework and making it easily accessible for physically distributed users (PaaS principle) will significantly reduce scientists' time in collecting and preprocessing data, which always takes over 80% of the scientific analysis. This will greatly speed up the knowledge discovery process. To ensure that the A2CI platform stays relevant, we will continue to enhance the system and make it more intelligent in terms of understanding users' search and analysis behavior in order to suggest most suitable data and tools to conduct online analysis. We will also closely collaborate with atmospheric scientists to integrate more simulation models and analysis tools to benefit a broader user community.

Acknowledgement

This work is supported in part by the National Science Foundation under Grants PLR-1349259, BCS-1455349 and PLR-1504432. Any opinions, findings, and conclusions or recommendations expressed in this material are those of the author and do not necessarily reflect the views of the National Science Foundation.

**References**

Eaton, B., Gregory, J., Drach, B., Taylor, K., Hankin, S., Caron, J., ... & Juckes, M. (2003). NetCDF Climate and Forecast (CF) metadata conventions. http://cfconventions.org/ (last accessed on September 11, 2015)

Fortner, B. (1998). HDF: The hierarchical data format. http://www.hdfgroup.org/HDF5/ (last accessed on September 11, 2015)

Fowler, D., Pilegaard, K., Sutton, M. A., Ambus, P., Raivonen, M., Duyzer, J., ... & Zechmeister-Boltenstern, S. (2009). Atmospheric composition change: ecosystems–atmosphere interactions. *Atmospheric Environment*, *43*(33), 5193-5267.

Guilyardi, E., Wittenberg, A., Fedorov, A., Collins, M., Wang, C., Capotondi, A., ... & Stockdale, T. (2009). Understanding El Niño in ocean-atmosphere general circulation models: Progress and challenges. *Bulletin of the American Meteorological Society*, 90(3), 325-340.

Katul, G. G., Oren, R., Manzoni, S., Higgins, C., & Parlange, M. B. (2012). Evapotranspiration: A process driving mass transport and energy exchange in the soil-plant-atmosphere-climate system. *Reviews of Geophysics*, *50*(3).

Karl, T. R., & Trenberth, K. E. (2003). Modern global climate change. *Science*, *302*(5651), 1719-1723.

Li, W., Yang, C., Nebert, D., Raskin, R., Houser, P., Wu, H., & Li, Z. (2011). Semantic-based web service discovery and chaining for building an Arctic spatial data infrastructure. *Computers & Geosciences*, *37*(11), 1752-1762.

Li, W., Goodchild, M. F., Anselin, L., & Weber, K. (2013). A service-oriented smart CyberGIS framework for data-intensive geospatial problems. *CyberGIS: Fostering a New Wave of Geospatial Discovery and Innovation*. (in press)

Li, W., Bhatia, V., & Cao, K. (2014). Intelligent polar cyberinfrastructure: enabling semantic search in geospatial metadata catalogue to support polar data discovery. *Earth Science Informatics*, *8*(1), 111-123.

Li, W., Song, M., Zhou, B., Cao, K., & Gao, S. (2015). Performance improvement techniques for geospatial web services in a cyberinfrastructure environment–A case study with a disaster management portal. *Computers, Environment and Urban Systems*. doi:10.1016/j.compenvurbsys.2015.04.003

Mattmann, C. A., Waliser, D., Kim, J., Goodale, C., Hart, A., Ramirez, P., ... & Hewitson, B. (2014). Cloud computing and virtualization within the regional climate model and evaluation system. *Earth Science Informatics*, *7*(1), 1-12.

Plale, B., Gannon, D., Brotzge, J., Droegemeier, K., Kurose, J., McLaughlin, D., ... & Chandrasekar, V. (2006). Casa and lead: Adaptive cyberinfrastructure for real-time multiscale weather forecasting. *IEEE Computer*, *39*(11), 56-64.

Ramanathan, V. C. P. J., Crutzen, P. J., Kiehl, J. T., & Rosenfeld, D. (2001). Aerosols, climate, and the hydrological cycle. *Science*, *294*(5549), 2119-2124.

Rew, R., & Davis, G. (1990). NetCDF: an interface for scientific data access. *Computer Graphics and Applications, IEEE*, *10*(4), 76-82.

Schneider, T. (2006). The general circulation of the atmosphere. *Annu. Rev. Earth Planet. Sci.*, *34*, 655-688.

Trenberth, K. E., Dai, A., van der Schrier, G., Jones, P. D., Barichivich, J., Briffa, K. R., & Sheffield, J. (2014). Global warming and changes in drought. *Nature Climate Change*, *4*(1), 17-22.

Walther, G. R., Post, E., Convey, P., Menzel, A., Parmesan, C., Beebee, T. J., ... & Bairlein, F. (2002). Ecological responses to recent climate change. *Nature*, *416*(6879), 389-395.

Wang, S. (2010). A CyberGIS framework for the synthesis of cyberinfrastructure, GIS, and spatial analysis. *Annals of the Association of American Geographers*, *100*(3), 535-557.

Wang, L., Von Laszewski, G., Younge, A., He, X., Kunze, M., Tao, J., & Fu, C. (2010). Cloud computing: a perspective study. *New Generation Computing*, *28*(2), 137-146.

Yang, C., Li, W., Xie, J., & Zhou, B. (2008). Distributed geospatial information processing: sharing distributed geospatial resources to support Digital Earth. *International Journal of Digital Earth*, *1*(3), 259-278.

Yang, C., Nebert, D., & Taylor, D. F. (2011). Establishing a sustainable and cross-boundary geospatial cyberinfrastructure to enable polar research. *Computers & Geosciences*, *37*(11), 1721-1726.

Yang, C., Raskin, R., Goodchild, M., & Gahegan, M. (2010). Geospatial cyberinfrastructure: past, present and future. *Computers, Environment and Urban Systems*, *34*(4), 264-277.